\documentclass[sigconf,nonacm]{acmart}
\AtBeginDocument{%
  }

\usepackage{multicol}
\usepackage{multirow}
\usepackage{booktabs}
\usepackage[table]{xcolor} 

\begin{document}

\title{Asymmetric Generative Recommendation via Kronecker Residual Bridge and Multi-Faceted Hierarchical Quantization}

\author{Bin Huang}
\affiliation{
  \institution{DCST, Tsinghua University}
  \country{Beijing, China}
}
\email{huangb23@mails.tsinghua.edu.cn}

\author{Xin Wang}
\authornote{Corresponding authors. DCST is the abbreviation of Department of Computer Science and Technology. BNRist is the abbreviation of Beijing National Research Center for Information Science and Technology.}
\affiliation{
  \institution{DCST, BNRist, Tsinghua University}
  \country{Beijing, China}
}
\email{xin_wang@tsinghua.edu.cn}

\author{Junwei Pan}
\affiliation{
  \institution{Tencent}
  \country{Shenzhen, China}
}
\email{jonaspan@tencent.com}

\author{Yongqi Zhou}
\affiliation{
  \institution{Tencent}
  \country{Shenzhen, China}
}
\email{kolinzhou@tencent.com}

\author{Yifeng Zhou}
\affiliation{
  \institution{Tencent}
  \country{Shenzhen, China}
}
\email{joefzhou@tencent.com}

\author{Zhixiang Feng}
\affiliation{
  \institution{Tencent}
  \country{Shenzhen, China}
}
\email{lionelfeng@tencent.com}

\author{Shudong Huang}
\affiliation{
  \institution{Tencent}
  \country{Shenzhen, China}
}
\email{ericdhuang@tencent.com}

\author{Haijie Gu}
\affiliation{
  \institution{Tencent}
  \country{Shenzhen, China}
}
\email{jerrickgu@tencent.com}

\author{Wenwu Zhu}
\authornotemark[1]
\affiliation{
  \institution{DCST, BNRist, Tsinghua University}
  \country{Beijing, China}
}
\email{wwzhu@tsinghua.edu.cn}

\renewcommand{\shortauthors}{Huang et al.}

\begin{abstract}
Generative Recommendation (GenRec) models reformulate recommendation as a sequence generation task, representing items as discrete Semantic IDs used symmetrically as both inputs and prediction targets. We identify a critical dual-stage information bottleneck in this design: (1) the Input Bottleneck, where lossy quantization degrades fine-grained semantics, while popularity bias skews the learned representations toward frequent items, and (2) the Output Bottleneck, where imprecise discrete targets limit supervision quality. To address these issues, we propose AsymRec, an asymmetric continuous-discrete framework that decouples input and output representations. Specifically, Kronecker Residual Bridge (KRB) maps continuous embeddings into the Transformer’s hidden space via a Kronecker projection with a residual pathway, preserving semantic richness and improving generalization to infrequent items. Multi-faceted Hierarchical Quantization (MHQ) constructs high-capacity, structured discrete targets through multi-view and multi-level quantization with semantic regularization, preventing dimensional collapse while retaining fine-grained distinctions. Extensive experiments demonstrate that AsymRec consistently outperforms state-of-the-art generative recommenders by an average of 18.7\%. \footnote{Our project page is at \url{https://github.com/huangb23/AsymRec}}
\end{abstract}


\maketitle

\begin{figure}[htbp]
  \centering
  \includegraphics[width=\linewidth]{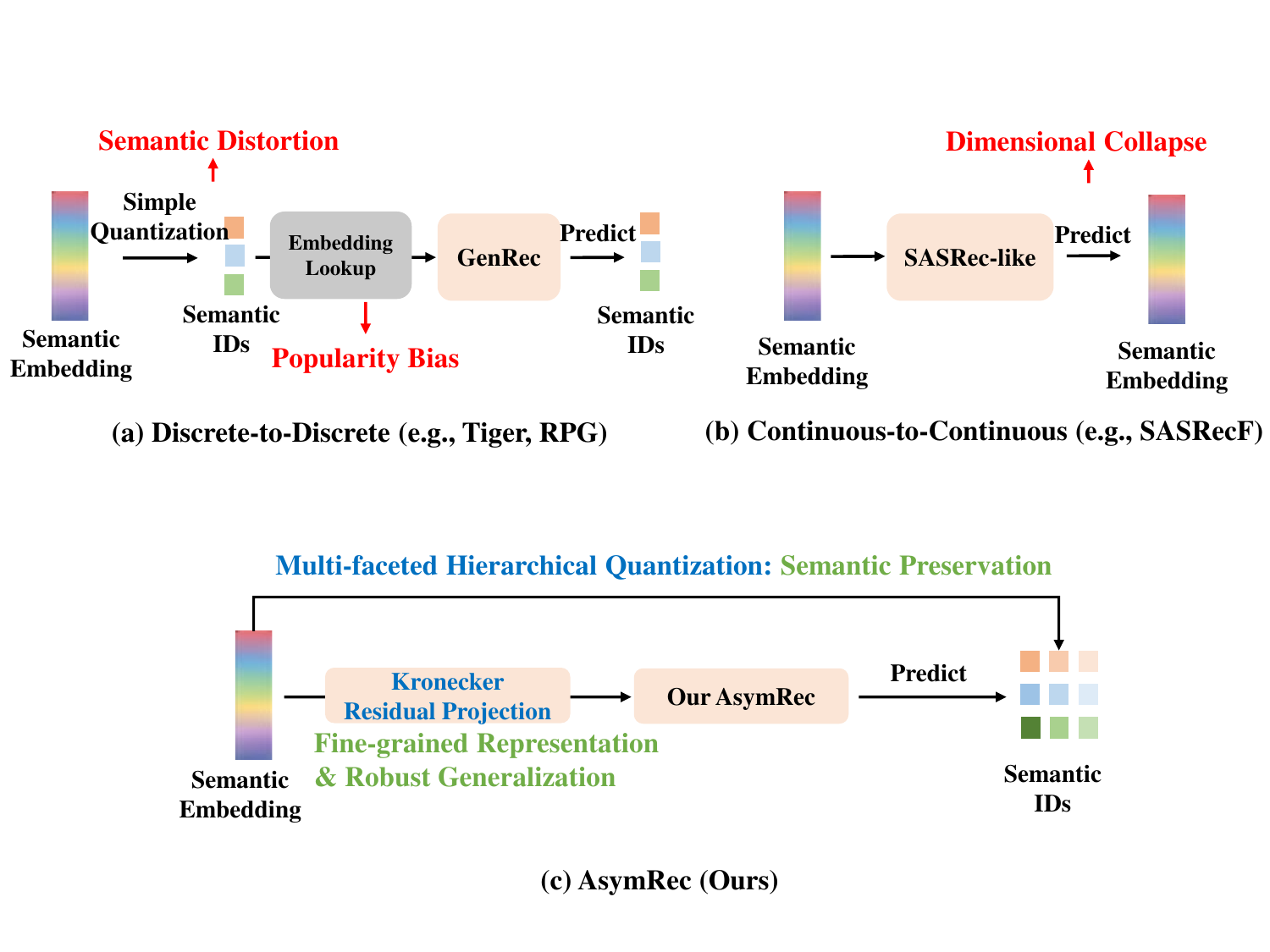}
  \caption{Comparison of different representation paradigms for recommendation. (a) Existing GenRec models adopt a discrete-to-discrete paradigm, using quantized item representations as both inputs and generation targets. (b) SASRec-like models follow a continuous-to-continuous paradigm by directly modeling continuous item embeddings. (c) Our method introduces an asymmetric continuous-to-discrete paradigm, decoupling input representation and output supervision via Kronecker residual projection and multi-faceted hierarchical quantization.}
  \label{fig:intro}
  \Description{Diagram showing the dual-stage information bottleneck in generative recommendation: symmetric quantization causes semantic distortion and popularity bias at the input, and imprecise supervision at the output. AsymRec decouples input and output via Kronecker residual projection and multi-faceted hierarchical quantization.}
\end{figure}

\section{Introduction}

Recent advancements have given rise to Generative Recommendation (GenRec) models, which reformulate recommendation as a sequence-to-sequence generation task~\cite{hstu_zhai2024actions,deng2025onerec,zhou2025onerecv2}.
Drawing inspiration from the success of large language models~\cite{yang2025qwen3, guo2025deepseek}, these approaches represent items as sequences of discrete tokens---commonly referred to as \emph{Semantic IDs}~\cite{rajput2023recommender,rpg_hou2025generating,vqrec_hou2023learning,petrov2024recjpq}.
This paradigm enables an auto-regressive Transformer to predict the next item in a unified generative manner~\cite{rajput2023recommender,hstu_zhai2024actions,rpg_hou2025generating}, offering the potential for better capture of long-range dependencies and seamless integration of multi-modal item features~\cite{vqrec_hou2023learning,agarwal2025pinrec}.

The efficacy of GenRec models hinges on the bridge between continuous item semantics and discrete generative tokens. Prevailing methods typically adopt a fully-discretized pipeline: they first quantize high-dimensional semantic embeddings (derived from text or visual encoders) into semantic IDs via RQ-VAE / vector-quantized autoencoding~\cite{lee2022autoregressive,oord2017neural,rajput2023recommender}, and then use these discrete IDs as both the input representation and the prediction target~\cite{rajput2023recommender,rpg_hou2025generating,vqrec_hou2023learning}. This symmetric design has become a common paradigm in recent GenRec systems, spanning both retrieval and ranking scenarios~\cite{rajput2023recommender,agarwal2025pinrec,huang2025towards,deng2025onerec,zhou2025onerecv2}.

However, this fully-discretized paradigm tightly couples the model performance with the quality of semantic quantization. As illustrated in Fig.~\ref{fig:intro} (a), existing GenRec models rely on the same quantized representation for both input encoding and output generation. While such a unified representation simplifies the generative process, it also introduces a Dual-stage Information Bottleneck, where the limitations of quantization affect both the learned item representations and the generated supervision signals.

\textbf{The Input Bottleneck: Semantic Distortion and Popularity Bias.} Traditional generative recommenders map discrete IDs into a learned embedding space via a lookup table before feeding them into the Transformer. This process introduces two key limitations. First, the initial quantization is inherently lossy, discarding fine-grained semantic nuances that cannot be recovered. Second, the learned ID embeddings tend to overemphasize frequently occurring “hot” items in the training set, limiting the model’s ability to generalize to less frequent “cold” items.

\textbf{The Output Bottleneck: Imprecise Supervision Signals.} On the output side, imperfect quantization produces noisy targets due to reconstruction errors and codebook collisions, limiting supervision quality. While one might consider predicting continuous embeddings directly to avoid this, such a transition often leads to "dimensional collapse,"~\cite{dimension_hua2021feature} where the model's output distribution shrinks to a narrow subspace, failing to distinguish between items with high precision. Thus, a high-capacity discrete target remains essential for effective supervision.

These observations suggest that the optimal representations for encoding user context and supervising generation are fundamentally different. The input stage benefits from preserving continuous semantic information, whereas the output stage still requires expressive discrete targets to provide stable and discriminative supervision. This motivates an asymmetric representation strategy that decouples the input and output representations.

To address these challenges, we propose AsymRec, a high-fidelity generative framework that decouples the input and output representations to maximize semantic preservation and improve generalization. To eliminate the input bottleneck, we introduce Kronecker Residual Bridge (KRB), which maps original continuous embeddings directly into the Transformer's hidden space via a Kronecker projection with a residual pathway, bypassing discrete ID lookup entirely. The residual pathway provides a stable linear baseline that preserves the original semantic geometry, enabling better generalization to less frequent “cold” items, while the Kronecker-structured correction captures fine-grained semantic adjustments with a compact, factorized parameterization. To address the output bottleneck, we propose Multi-faceted Hierarchical Quantization (MHQ). MHQ first applies a learnable projection to reorganize embeddings into a structured latent space, explicitly regularized to balance semantic energy across subspaces and to reduce redundancy and correlation among them. Built upon this representation, MHQ performs hierarchical, multi-path quantization with an Exponential Moving Average (EMA) strategy to stabilize the discrete optimization process. This results in a multi-dimensional, multi-layer coordinate system that yields high-capacity discrete targets, effectively preventing dimensional collapse while retaining fine-grained semantic distinctions.

Our key contributions are summarized as follows:
\begin{itemize}
    \item We identify and analyze the \textbf{dual-stage information bottleneck} in generative recommendation, highlighting how discrete inputs bias the model toward frequently occurring “hot” items and how standard quantized outputs limit prediction precision.
    \item We propose \textbf{Kronecker Residual Bridge (KRB)}, which replaces traditional ID lookup with a continuous, Kronecker-structured projection, preserving fine-grained semantic topology to enhance generalization to less frequent “cold” items and improve prediction accuracy.
    \item We develop \textbf{Multi-faceted Hierarchical Quantization (MHQ)}, a structured discretization framework that integrates learnable projection, structural regularization, and EMA-stabilized hierarchical quantization to provide high-fidelity discrete supervision while maintaining semantic and hierarchical consistency.
    \item Extensive experiments on the Amazon public benchmarks demonstrate that \textbf{AsymRec} consistently outperforms state-of-the-art generative recommenders by an average of \textbf{18.7\%}. Beyond offline evaluation, we further deploy \textbf{AsymRec} in a production pCVR system on one of the world’s largest advertising platforms, where it achieves a 1.4\% lift in total consumption and a 1.9\% GMV uplift in online A/B tests, validating its effectiveness at industrial scale.
\end{itemize}

\section{Related Works}

\subsection{Generative Recommendation}

Sequential recommendation (SR) aims to capture the dynamic evolution of user preferences by modeling historical interaction sequences. Traditional discriminative approaches, from early Markov Chains~\cite{rendle2010factorizing} to modern Transformer-based models like SASRec~\cite{sasrec_kang2018self} and BERT4Rec~\cite{sun2019bert4rec}, primarily frame recommendation as a ranking task. They operate by scoring items from a fixed corpus, treating item IDs as independent, atomic tokens. This paradigm faces intrinsic challenges: it struggles with cold-start scenarios~\cite{cold1_zhu2021learning, cold2_xu2024cmclrec} due to the sparsity and lack of inherent meaning in random IDs, and it fails to explicitly leverage the rich semantic correlations between items, limiting generalization.

To overcome these limitations, Generative Recommendation (GR)~\cite{rajput2023recommender, genrec_li2025survey} has recently emerged as a promising alternative paradigm. Instead of relying on learned item ID embeddings, GR models first encode item-side semantic content—such as titles, descriptions, or other multi-modal attributes—into continuous semantic embeddings. These embeddings are then discretized into a set of semantic IDs that capture high-level semantic attributes of items. Given a user’s interaction history, the GR model takes the semantic IDs of previously interacted items as input and autoregressively generates the semantic IDs corresponding to the target item for recommendation.

\begin{figure*}[tbp]
  \centering
  \includegraphics[width=0.95\textwidth]{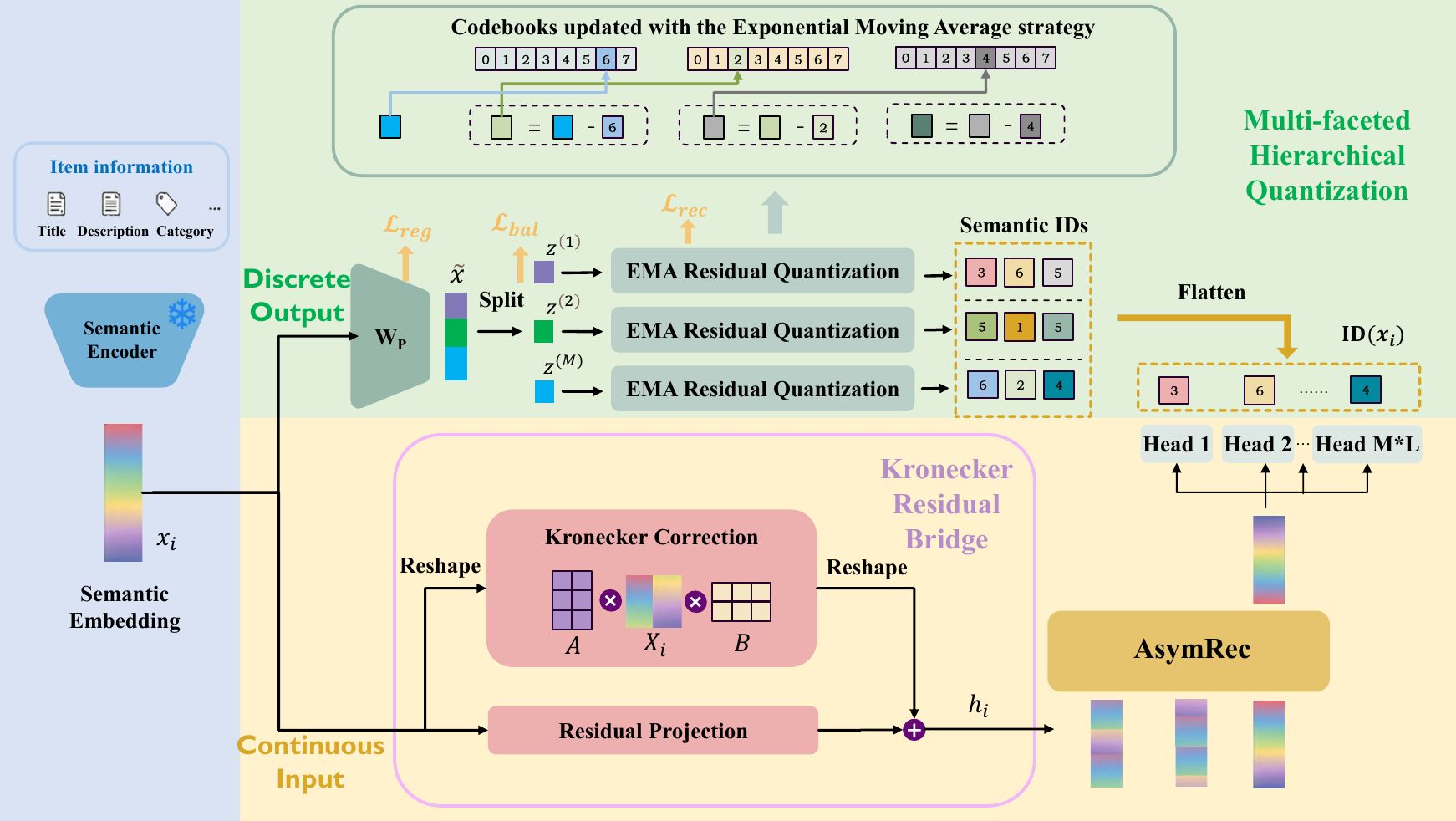}
  \caption{
Overview of the proposed \textbf{AsymRec} framework. 
The input item is first encoded into a \textbf{continuous semantic embedding}, which is mapped by the \textbf{Kronecker Residual Bridge (KRB)} module to produce fine-grained representations. 
These representations are then fed into the \textbf{Transformer Decoder} to predict the corresponding \textbf{semantic IDs} generated by the \textbf{Multi-faceted Hierarchical Quantization (MHQ)} module. 
MHQ maps each embedding into multiple subspaces and applies EMA Residual Quantization in each subspace, producing fine-grained discrete semantic IDs. 
}

  \label{fig:framework}
  \Description{Overview of the AsymRec framework. A continuous semantic embedding is mapped by the KRB module into fine-grained representations, which are fed into a Transformer Decoder. The MHQ module maps each embedding into multiple subspaces and applies EMA residual quantization to produce discrete semantic IDs.}
\end{figure*}

\subsection{Semantic ID Generation}
\label{sec:sid_related}
A key component of Generative Recommendation is the discretization of continuous semantic embeddings into semantic IDs (SIDs), which enables item representation and generation in a discrete token space. Existing methods for SID generation can be broadly categorized into residual quantization (RQ)–based methods and product quantization (PQ)–based methods, each offering distinct advantages and limitations.

Residual quantization~\cite{rq1_juang1982multiple, rq2_martinez2014stacked} is the most widely adopted approach for semantic ID generation. By iteratively quantizing the residual between the original embedding and previously selected codewords, RQ constructs a hierarchical, coarse-to-fine representation. This multi-level structure aligns naturally with autoregressive generation, as it progressively narrows the candidate search space and captures semantic information at different levels of granularity. RQ-based methods have been extensively applied in generative recommender systems~\cite{rajput2023recommender, deng2025onerec}. However, a significant drawback of RQ is its tendency toward \textit{semantic entanglement}. Because all residual levels are optimized along a single path, RQ often struggles to disentangle independent semantic facets (e.g., brand, category, and style), potentially conflating distinct item attributes into a coupled ID sequence.

Product quantization~\cite{pq_gray1984vector} decomposes the embedding space into multiple independent subspaces and quantizes each separately, enabling fine-grained modeling of different semantic aspects~\cite{rpg_hou2025generating}. While PQ effectively captures multi-faceted information, it lacks the \textit{hierarchical depth} inherent in RQ.

\section{Method}

In this section, we introduce our model, AsymRec, as illustrated in Fig.~\ref{fig:framework}. We first formally define the generative recommendation problem and describe the representation of items in both continuous and discrete spaces. Then, we present the two core components of our framework: Kronecker Residual Bridge (KRB), which preserves rich semantic information by mapping continuous embeddings into the model’s feature space, and Multi-faceted Hierarchical Quantization (MHQ), which produces high-fidelity, structured discrete targets for supervision. Finally, we describe the overall architecture of AsymRec, which integrates KRB and MHQ within a Transformer-based generative recommendation model to effectively capture fine-grained user preferences and generate coherent recommendations.

\subsection{Problem Definition}

Sequential recommendation aims to model user preferences based on historical interactions and predict the next item a user is likely to interact with. Formally, given a user interaction sequence
\[
\mathcal{S}_u = [I_1, I_2, \dots, I_{T}],
\]
where \(I_i\) denotes the \(i\)-th item, the task is to predict the next item \(I_{T+1}\).

In the generative recommendation setting, each item is represented by a continuous semantic embedding $x_i \in \mathbb{R}^{d}$. To enable generation in a discrete space, each embedding is quantized into a set of semantic IDs, denoted as $\mathbf{ID}(x_i)$. The task objective is then to predict the semantic IDs of the next item, denoted as $\mathbf{ID}(x_{T+1})$, based on the preceding items in the sequence.

\subsection{Kronecker Residual Bridge (KRB)}

For a user interaction sequence $\mathcal{S}_u $,
each item is represented by a continuous semantic embedding $x_i$ as well as multiple quantized tokens $\mathbf{ID}(x_i)$. 
In prior generative recommendation methods, only the discrete token embeddings are used as input: 
the tokens are typically mapped to embeddings via a lookup table and then either concatenated  along the sequence dimension~\cite{rajput2023recommender} or averaged~\cite{rpg_hou2025generating} to form the input to the recommendation model, 
while the original continuous embedding $x_i$ is never utilized.

We identify two limitations of this approach: 
(1) \emph{quantization is inherently lossy}, discarding fine-grained semantic nuances that cannot be recovered, which prevents the model from distinguishing cold or subtly different items; 
(2) \emph{learning bias toward popular items}, as embeddings of frequent IDs are updated far more often during training, while rare items receive insufficient supervision and remain under-trained.

To address these issues, a natural baseline is to project the continuous embedding $x_i$ into the recommendation space via a multi-layer MLP. Our preliminary experiments confirm that even a simple MLP projector significantly outperforms the discrete-input baseline (see Row 3 in Table~\ref{tab:ablation}), validating the value of continuous inputs. However, we observe that the large number of unconstrained parameters in an MLP tends to overfit to training-item-specific patterns, distorting the semantic geometry inherited from the pretrained encoder.

To overcome this, we propose \textbf{Kronecker Residual Bridge (KRB)}, which decouples the projection into two complementary branches: a residual linear projection that preserves the pretrained semantic geometry, and a compact Kronecker-structured correction that injects additional expressive capacity with minimal risk of overfitting. The key insight is that the Kronecker form leverages the rank multiplication property $\text{rank}(A \otimes B) = \text{rank}(A) \times \text{rank}(B)$: despite having only $n_1 m_1 + n_2 m_2$ parameters (e.g., 2560 in our configuration), the correction term can achieve full-rank expressiveness up to $n_1 n_2 = d_m$. This allows KRB to match or exceed the capacity of an MLP projector while preserving the structural properties of the semantic embedding space.

Specifically, given an input item embedding $x_i \in \mathbb{R}^{d}$, 
KRB maps it into the recommendation model's feature space:
\begin{equation}
    h_i = \text{KRB}(x_i) \in \mathbb{R}^{d_m}.
\end{equation}

Concretely, a residual bridge projects $x_i$ into the recommendation latent space:
\begin{equation}
    h_i^{r}=W_r\,x_i \in \mathbb{R}^{d_m},
\end{equation}
where $W_r\in\mathbb{R}^{d_m\times d}$ is a learnable linear projection. 
This residual pathway provides a direct information route for adapting pretrained semantic representations to the recommendation feature space.

To further capture structured semantic transformations beyond the residual projection, KRB introduces a Kronecker-structured correction branch. Specifically, the representation is reshaped as:
\begin{equation}
    X_i=\operatorname{reshape}(x_i)\in\mathbb{R}^{m_1\times m_2},
\end{equation}
where $d=m_1 m_2$. The correction term is parameterized by two learnable factors:
\begin{equation}
    A\in\mathbb{R}^{n_1\times m_1}, \quad
    B\in\mathbb{R}^{n_2\times m_2},
\end{equation}
with $n_1 n_2 = d_m$, and computed as:
\begin{equation}
    h_i^{k}=\operatorname{reshape}(A X_i B^{\top}) \in \mathbb{R}^{d_m},
\end{equation}
which is equivalent to:
\begin{equation}
    h_i^{k}=(A\otimes B)\,x_i.
\end{equation}

The final representation is obtained by combining the residual bridge and the Kronecker correction:
\begin{equation}
    h_i = h_i^{r} + h_i^{k}
    =
    W_r\,x_i
    + (A\otimes B)\,x_i.
\end{equation}

This dual-branch design enables KRB to efficiently project high-dimensional semantic embeddings ($d=3072$) to the model dimension ($d_m=448$) while preserving semantic geometry. The residual bridge guarantees a stable baseline mapping, while the Kronecker-structured correction captures fine-grained, multi-dimensional semantic adjustments. This representation serves as the input to downstream generative recommendation modules.

\subsection{Multi-faceted Hierarchical Quantization (MHQ)}

With the input-side quantization loss addressed by KRB, we now turn to the output side.
While predicting continuous vectors at the output could seem natural, this often causes dimensional collapse (Sec.~\ref{sec:continuous_output}), resulting in suboptimal performance. 
Instead, predicting discrete semantic identifiers preserves the structure and improves generative recommendation quality. 

Nevertheless, as pointed out in Sec.~\ref{sec:sid_related}, existing quantization methods are insufficient to capture both multi-faceted and hierarchical semantic information simultaneously. 
To address this, we propose the \textbf{Multi-faceted Hierarchical Quantization (MHQ)} module. 
MHQ combines the strengths of residual quantization (RQ) and product quantization (PQ) by first partitioning the embedding into multiple orthogonal subspaces, 
and then applying hierarchical residual quantization within each subspace. 
This design ensures that the generated semantic IDs are both multi-dimensional in semantic coverage and progressively detailed within each dimension.

Given a semantic embedding vector $x \in \mathbb{R}^d$, the MHQ module first projects it into a latent space $\mathbb{R}^{D}$ via a learnable linear transformation $\tilde{x} = W_P x$, where $W_P \in \mathbb{R}^{D \times d}$. To extract diverse semantic facets, the projected vector $\tilde{x}$ is partitioned into $M$ disjoint subspaces:
\begin{equation}
\tilde{x} = [z^{(1)}, z^{(2)}, \dots, z^{(M)}], \quad z^{(m)} \in \mathbb{R}^{d_s}
\end{equation}
where $d_s = D/M$ denotes the dimensionality of each subspace. 

Within each subspace $m$, we implement a Residual Quantization process of depth $L$. For each level $l \in \{1, \dots, L\}$, a codebook $\mathcal{C}^{(m,l)} = \{c_k^{(m,l)}\}_{k=1}^K$ is maintained, where $K$ is the codebook size. The quantizer iteratively identifies the optimal centroid index $i_{m,l}$ by minimizing the $L_2$ distance between the current residual $r_l^{(m)}$ and the codebook entries:
\begin{equation}
i_{m,l} = \arg\min_{k \in \{1, \dots, K\}} \| r_l^{(m)} - c_k^{(m,l)} \|_2^2
\end{equation}
where $r_1^{(m)} = z^{(m)}$, and the residual for the subsequent level is updated as $r_{l+1}^{(m)} = r_l^{(m)} - c_{i_{m,l}}^{(m,l)}$. The reconstructed representation in the $m$-th subspace is thus $\hat{z}^{(m)} = \sum_{l=1}^L c_{i_{m,l}}^{(m,l)}$.

To stabilize the discrete optimization, we employ the Exponential Moving Average (EMA) strategy for codebook updates instead of standard backpropagation. For a given centroid $c_k^{(m,l)}$, the update rules are:
\begin{align}
N_{k}^{(m,l)} &\leftarrow \gamma N_{k}^{(m,l)} + (1 - \gamma) \sum_{j=1}^B \mathbb{1}[i_{m,l}^{(j)} = k] \\
m_{k}^{(m,l)} &\leftarrow \gamma m_{k}^{(m,l)} + (1 - \gamma) \sum_{j=1}^B \mathbb{1}[i_{m,l}^{(j)} = k] r_l^{(m,j)} \\
c_k^{(m,l)} &= \frac{m_k^{(m,l)}}{N_k^{(m,l)}}
\end{align}
where $\gamma$ is the decay factor and $B$ is the batch size. Finally, each item is represented by a flattened sequence of indices $\mathbf{ID}(x) = \{i_{1,1}, i_{1,2}, i_{1,3}, \dots, i_{M,L}\}$, forming a structured semantic codeword of length $M \times L$.

The training objective of MHQ is formulated as a multi-task loss function. The primary component is the reconstruction loss:
\begin{equation}
    \mathcal{L}_{rec} = \| \tilde{x} - \text{concat}(\hat{z}^{(1)}, \dots, \hat{z}^{(M)}) \|_2^2, 
\end{equation}

which ensures high fidelity of the quantized IDs. To prevent information from collapsing into a subset of subspaces, we introduce a subspace energy balance loss that explicitly penalizes uneven energy allocation across different facets. Specifically, let $\mathbb{E}[\|z^{(m)}\|_2^2]$ denote the expected energy of the $m$-th subspace. We first compute the mean energy across all $M$ subspaces:
\begin{equation}
    \bar{E} = \frac{1}{M} \sum_{m=1}^M \mathbb{E}\big[\|z^{(m)}\|_2^2\big],
\end{equation}
and define the balance loss as the mean absolute deviation from this average:
\begin{equation}
    \mathcal{L}_{bal} = \frac{1}{M} \sum_{m=1}^M \left| \mathbb{E}\big[\|z^{(m)}\|_2^2\big] - \bar{E} \right|.
\end{equation}
This formulation encourages an equitable distribution of information across all $M$ facets.

In addition, to reduce redundancy and correlation among different subspaces, we impose an orthogonality regularization on the projection matrix $W_P$, defined as
\begin{equation}
    \mathcal{L}_{reg} = \left\| W_P W_P^\top - I \right\|_F,
\end{equation}
where $I$ denotes the identity matrix.

The overall training objective is given by
\begin{equation}
    \mathcal{L}_{MHQ} = \mathcal{L}_{rec} + \lambda_{bal}\mathcal{L}_{bal} + \lambda_{reg}\mathcal{L}_{reg}.
\end{equation}

This loss is applied only during the training of MHQ and is not used in the subsequent training of the recommendation model. After training, the discrete tokens $\textbf{ID}(x_i)$ are assigned to each $x_i$ based on the learned quantization.

\subsection{AsymRec~Architecture}

In this section, we present the overall architecture of AsymRec, as illustrated in Fig.~\ref{fig:framework}. The framework adopts an asymmetric continuous-discrete design: continuous embeddings are mapped into the model’s feature space via KRB, while high-fidelity, multi-faceted discrete targets are produced by MHQ for supervision. These two complementary components are integrated within a Transformer-based generative model.

Given a sequence of item embeddings corresponding to a user's interactions $[x_1, x_2, \dots, x_T]$
each item embedding $x_i$ is first mapped into the recommendation feature space via the Kronecker Residual Bridge (KRB) module as $h_i = \text{KRB}(x_i)$. 

Positional encodings are then added:
\begin{equation}
    \mathbf{H}^{0} = [h_1 + p_1, h_2 + p_2, \dots, h_T + p_T].
\end{equation}
The resulting sequence $\mathbf{H}^{0}$ is then fed into $L_T$ Transformer decoder layers. Each layer utilizes multi-head self-attention and feed-forward networks to model the complex transitions between user interests:

\begin{equation}
H^i = \text{Decoder}(H^{i-1}), \quad i = 1, \dots, L_T,
\end{equation}

We extract the hidden state of the last item from the final decoder layer, denoted as $\mathbf{H}^{L_T}_{T} \in \mathbb{R}^{d_m}$, and feed it into $M \times L$ parallel prediction heads to predict the structured semantic IDs of the next item, $\textbf{ID}(x_{T+1})$. 
Each prediction head is implemented as a two-layer MLP that maps $\mathbf{H}^{L_T}_{T}$ to a $K$-way categorical distribution over the corresponding quantized codebook entries. 
Formally, we optimize the cross-entropy loss over all heads:
\begin{equation}
    \mathcal{L}_{\text{CE}} = - \frac{1}{ML}  \sum_{m=1}^M \sum_{l=1}^L \log p\big(i_{m,l}^{T+1} \,\big|\, \text{model}(x_{\le T})\big),
\end{equation}
which encourages accurate semantic ID predictions across all facets and hierarchical layers. 
At inference, we employ a graph-constrained decoding strategy~\cite{rpg_hou2025generating} to ensure that only valid codewords are generated.

This asymmetric design naturally integrates the continuous input mapping from KRB with the multi-faceted, hierarchical quantization of MHQ, 
allowing AsymRec to capture fine-grained item semantics while producing coherent and structured recommendations.

\begin{table}[tbp] 
  \small
  \centering
\caption{Statistics of the processed datasets. ``Avg.~$t$'' denotes the average number of interactions per input sequence.}
\label{tab:dataset}
\begin{tabular}{crrrr}
  \toprule
  \textbf{Datasets} & \textbf{\#Users} & \textbf{\#Items} & \textbf{\#Interactions} & \textbf{Avg.~$t$}\\
  \midrule
  \textbf{Sports}  & 18,357            & 35,598           & 260,739            & 8.32 \\
  \textbf{Beauty}  & 22,363            & 12,101           & 176,139            & 8.87 \\
  \textbf{Toys}    & 19,412            & 11,924           & 148,185            & 8.63 \\
  \textbf{CDs}     & 75,258            & 64,443           & 1,022,334          & 14.58 \\
  \bottomrule
\end{tabular}
\end{table}

\begin{table*}[t!]
  \small
    \centering
    \caption{Performance comparison among baselines and AsymRec. The best performance score is denoted in \textbf{bold}. The second-best performance score is denoted in \underline{underline}.
  }
    \label{tab:overall}
  \setlength{\tabcolsep}{1.1mm}{
    \begin{tabular}{@{}lcccccccccccccccc@{}}
      \toprule
      \multicolumn{1}{c}{\multirow{2}{*}{\textbf{Model}}} & \multicolumn{4}{c}{\textbf{Sports and Outdoors}} & \multicolumn{4}{c}{\textbf{Beauty}} & \multicolumn{4}{c}{\textbf{Toys and Games}} & \multicolumn{4}{c}{\textbf{CDs and Vinyl}} \\
      \cmidrule(l){2-5} \cmidrule(l){6-9}\cmidrule(l){10-13}\cmidrule(l){14-17}
      & R@5 & N@5 & R@10 & N@10 & R@5 & N@5 & R@10 & N@10 & R@5 & N@5 & R@10 & N@10 & R@5 & N@5 & R@10 & N@10 \\
      \midrule
      \multicolumn{17}{@{}c}{\textit{Item ID-based}} \\
      \midrule
      Caser~\cite{caser_tang2018personalized} & 0.0116 & 0.0072 & 0.0194 & 0.0097 & 0.0205 & 0.0131 & 0.0347 & 0.0176 & 0.0166 & 0.0107 & 0.0270 & 0.0141 & 0.0116 & 0.0073 & 0.0205 & 0.0101 \\
GRU4Rec~\cite{gru4rec_hidasi2015session} & 0.0129 & 0.0086 & 0.0204 & 0.0110 & 0.0164 & 0.0099 & 0.0283 & 0.0137 & 0.0097 & 0.0059 & 0.0176 & 0.0084 & 0.0195 & 0.0120 & 0.0353 & 0.0171 \\
HGN~\cite{hgn_ma2019hierarchical} & 0.0189 & 0.0120 & 0.0313 & 0.0159 & 0.0325 & 0.0206 & 0.0512 & 0.0266 & 0.0321 & 0.0221 & 0.0497 & 0.0277 & 0.0259 & 0.0153 & 0.0467 & 0.0220 \\
BERT4Rec~\cite{sun2019bert4rec} & 0.0115 & 0.0075 & 0.0191 & 0.0099 & 0.0203 & 0.0124 & 0.0347 & 0.0170 & 0.0116 & 0.0071 & 0.0203 & 0.0099 & 0.0326 & 0.0201 & 0.0547 & 0.0271 \\
SASRec~\cite{sasrec_kang2018self} & 0.0233 & 0.0154 & 0.0350 & 0.0192 & 0.0387 & 0.0249 & 0.0605 & 0.0318 & 0.0463 & 0.0306 & 0.0675 & 0.0374 & 0.0351 & 0.0177 & 0.0619 & 0.0263 \\
FDSA~\cite{fdsa_zhang2019feature} & 0.0182 & 0.0122 & 0.0288 & 0.0156 & 0.0267 & 0.0163 & 0.0407 & 0.0208 & 0.0228 & 0.0140 & 0.0381 & 0.0189 & 0.0226 & 0.0137 & 0.0378 & 0.0186 \\
S$^3$-Rec~\cite{zhou2020s3} & 0.0251 & 0.0161 & 0.0385 & 0.0204 & 0.0387 & 0.0244 & 0.0647 & 0.0327 & 0.0443 & 0.0294 & 0.0700 & 0.0376 & 0.0213 & 0.0130 & 0.0375 & 0.0182 \\
      \midrule
      \multicolumn{17}{@{}c}{\textit{Semantic ID-based}} \\
      \midrule
      RecJPQ~\cite{petrov2024recjpq} & 0.0141 & 0.0076 & 0.0220 & 0.0102 & 0.0311 & 0.0167 & 0.0482 & 0.0222 & 0.0331 & 0.0182 & 0.0484 & 0.0231 & 0.0075 & 0.0046 & 0.0138 & 0.0066 \\
VQ-Rec~\cite{vqrec_hou2023learning} & 0.0208 & 0.0144 & 0.0300 & 0.0173 & 0.0457 & 0.0317 & 0.0664 & 0.0383 & 0.0497 & 0.0346 & 0.0737 & 0.0423 & 0.0352 & 0.0238 & 0.0520 & 0.0292 \\
TIGER~\cite{rajput2023recommender} & 0.0264 & 0.0181 & 0.0400 & 0.0225 & 0.0454 & 0.0321 & 0.0648 & 0.0384 & 0.0521 & 0.0371 & 0.0712 & 0.0432 & 0.0492 & 0.0329 & \underline{0.0748} & 0.0411 \\
HSTU~\cite{hstu_zhai2024actions} & 0.0258 & 0.0165 & 0.0414 & 0.0215 & 0.0469 & 0.0314 & 0.0704 & 0.0389 & 0.0433 & 0.0281 & 0.0669 & 0.0357 & 0.0417 & 0.0275 & 0.0638 & 0.0346 \\    
RPG~\cite{rpg_hou2025generating} & \underline{0.0314} & \underline{0.0216} & \underline{0.0463} & \underline{0.0263} & \underline{0.0550} & \underline{0.0381} & \underline{0.0809} & \underline{0.0464} & \underline{0.0592} & \underline{0.0401} & \underline{0.0869} & \underline{0.0490} & \underline{0.0498} & \underline{0.0338} & 0.0735 & \underline{0.0415} \\ 
\midrule
\textbf{AsymRec} 
& \textbf{0.0369} & \textbf{0.0252} & \textbf{0.0553} & \textbf{0.0311}
& \textbf{0.0626} & \textbf{0.0446} & \textbf{0.0901} & \textbf{0.0535}
& \textbf{0.0693} & \textbf{0.0496} & \textbf{0.0955} & \textbf{0.0580}
& \textbf{0.0608} & \textbf{0.0415} & \textbf{0.0908} & \textbf{0.0511}     \\
      \bottomrule
    \end{tabular}
    }
  \end{table*}

\section{Experiment}

To evaluate the effectiveness of AsymRec and validate our hypotheses regarding the dual-stage bottleneck, we aim to answer the following research questions:

\begin{itemize} 
\item \textbf{RQ1: Overall Performance.} How does AsymRec perform compared to state-of-the-art generative and sequential recommendation baselines across various benchmarks?

\item \textbf{RQ2: Impact of Continuous Input on Optimization and Generalization.} Does the Kronecker Residual Bridge truly alleviate the input bottleneck? Specifically, how does it affect the model's ability to maintain semantic topology and generalize to cold items compared to discrete ID inputs?

\item \textbf{RQ3: Necessity of Discrete Output vs. Continuous Output.} Why not adopt a fully continuous pipeline? What are the empirical consequences when using continuous embeddings as the generation target?

\item \textbf{RQ4: Effectiveness of Multi-faceted Hierarchical Quantization (MHQ).} Does MHQ provide a higher-fidelity supervision signal than traditional quantization methods? How does the subspace decomposition affect recommendation precision?

\end{itemize}

\subsection{Experimental Setup}

\textbf{Dataset.} We conduct experiments on four widely-used categories from the Amazon Review benchmark~\cite{amazon_mcauley2015image}: \textbf{Sports and Outdoors} (\textbf{Sports}), \textbf{Beauty}, \textbf{Toys and Games} (\textbf{Toys}), and \textbf{CDs and Vinyl} (\textbf{CDs}). Following previous studies~\cite{rajput2023recommender,sasrec_kang2018self,zhou2020s3, rpg_hou2025generating}, we treat user reviews as interactions and organize them chronologically to construct interaction sequences. We follow the standard ``5-core'' filtering, ensuring each user and item has at least five interactions. The detailed statistics of the datasets are summarized in Table~\ref{tab:dataset}.

\textbf{Evaluation Protocol.} We adopt the widely used leave-last-out evaluation protocol, reserving the last item in each sequence for testing, the second-to-last item for validation, and the remaining prefix for training. To measure recommendation performance, we employ two widely-adopted ranking metrics: Recall@$K$ and Normalized Discounted Cumulative Gain (NDCG@$K$), with $K \in \{5, 10\}$. We report the test performance corresponding to the best results on the validation set.

\textbf{Implementation details.}  We use OpenAI's text-embedding-3-large as the semantic encoder following \citet{rpg_hou2025generating}, with output dimension $d = 3072$.  When training MHQ, we set the quantized embedding dimension $D = 512$, the loss weight $\lambda_{bal} = 0.01$, $\lambda_{reg} = 0.01$, the decay factor $\gamma = 0.99$, the learning rate to $0.001$, and train for $50$ epochs. For training AsymRec, we employ a Transformer decoder with $L_T = 2$ layers and an embedding dimension $d_m = 448$. For KRB, we set $m_1=48$, $m_2=64$ (such that $d = m_1 m_2 = 3072$), and $n_1=16$, $n_2=28$ (such that $n_1 n_2 = d_m = 448$). We train for a maximum of $100$ epochs with a batch size of $256$ and a learning rate of $0.003$. An early stopping strategy is adopted to halt training if validation performance does not improve for $20$ consecutive epochs. We tune the number of codebooks $M \in \{16, 32, 64\}$, the number of layers $L \in \{2, 3\}$, and the codebook size $K \in \{256, 512, 1024\}$. On the Beauty dataset, among the $12,101$ items, $12,099$ have unique codes; therefore, no additional collision handling is applied. Training and evaluation on the Beauty dataset complete within one hour using an NVIDIA GeForce RTX 3090 GPU.

\subsection{Overall Performance (RQ1)}
We compare AsymRec\ with item ID-based and semantic ID-based baselines across four datasets. The results are shown in Table~\ref{tab:overall}.

Compared to all baselines, the proposed AsymRec\ achieves the best overall performance, ranking first in all metrics. It outperforms the strongest baseline by an average of $18.7\%$ on the NDCG@10 metric.

\subsection{Ablation Study} 

\begin{table}[t]
\centering
\caption{Ablation study on the Beauty dataset.}
\label{tab:ablation}
\begin{tabular}{c l c}
\toprule
\textbf{Row} & \textbf{Variant} & \textbf{N@10} \\
\midrule
1 & \textbf{AsymRec} & \textbf{0.0535} \\
\midrule
2 & w/ discrete codes as inputs & 0.0491 \\
3 & w/ optimal MLP projector (3072→512→448) & 0.0508 \\
4 & w/ continuous embeddings as outputs & 0.0406 \\
5 & w/o MHQ & 0.0516 \\
\bottomrule
\end{tabular}
\end{table}

We conduct a series of ablation experiments to validate the effectiveness of the design choices in AsymRec. The primary results are summarized in Table~\ref{tab:ablation}.

\subsubsection{Impact of Input (RQ2)}

To investigate the necessity of the proposed \emph{Kronecker Residual Bridge (KRB)}, we compare AsymRec~(Row 1) with a variant that directly uses discrete semantic IDs (SIDs) as model inputs (Row 2). 
Specifically, in this variant, each SID token is mapped to a learnable embedding via a lookup table, and the embeddings of all tokens corresponding to an item are averaged to form its input representation~\cite{rpg_hou2025generating}:
\[
h_i = \frac{1}{|\mathbf{ID}(x_i) |} \sum_{i_{m,l} \in \mathbf{ID}(x_i)} \text{Emb}(i_{m,l})
\]
As shown in Table~\ref{tab:ablation}, replacing continuous embeddings with discrete codes leads to a performance drop.

To further isolate and analyze the impact of input representations, we conduct a frequency-aware similarity analysis at the representation level. 
For each user sequence, we compute the mean-pooled input representation of the historical items, $\bar{h} = \frac{1}{T} \sum_{i=1}^{T} h_i$, and measure its similarity to the representation of the ground-truth next item $h_{T+1}$ as well as 99 randomly sampled negative items. 
Based on these similarity scores, we compute Recall@10 across different item frequency bins.

\begin{figure}[tbp]
  \centering
  \includegraphics[width=\linewidth]{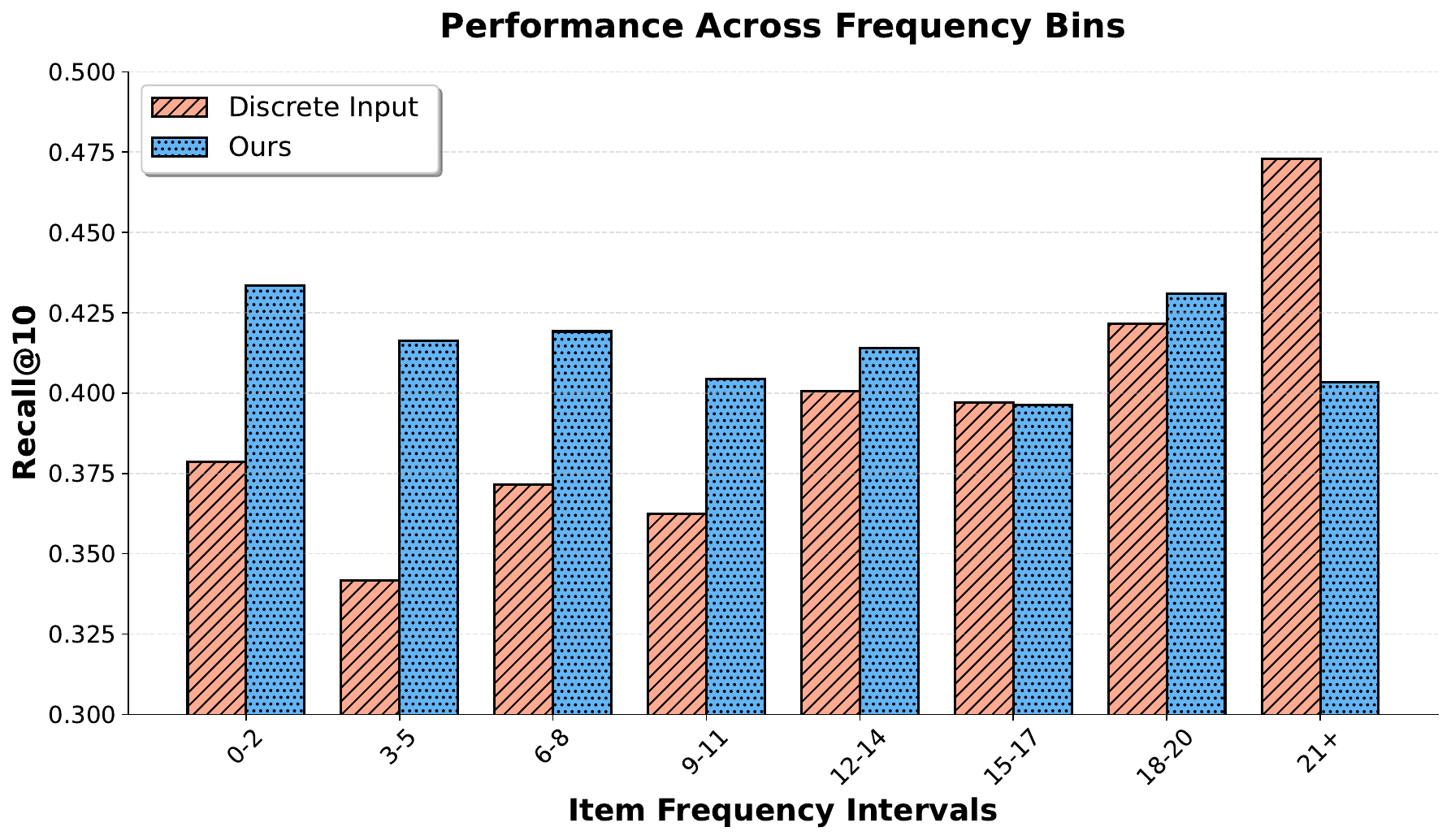}
  \caption{Retrieval performance at the input stage using Mean Pooling. Results are based on a 1-of-100 sampled ranking (1 positive target vs. 99 random negatives). 40\% of the items have a frequency of 6 or less, while 80\% of the items interact no more than 15 times.}
  \label{fig:bin_recall}
  \Description{Bar chart comparing Recall@10 across item frequency bins for discrete SID inputs versus AsymRec continuous inputs, showing AsymRec's advantage on low- and mid-frequency items.}
\end{figure}

The results are shown in Fig.~\ref{fig:bin_recall}. 
We observe that models using discrete SID inputs suffer from substantial performance degradation on low-frequency (cold) items, indicating poor generalization beyond popular items. 
In contrast, our model consistently achieves higher Recall@10 across almost all frequency bins, with particularly significant gains in the low- and mid-frequency regimes. 
This suggests that directly learning continuous input representations via KRB better preserves the topological structure of the original embedding space, enabling the model to capture fine-grained semantic relationships even for infrequent items. 
Notably, while the discrete-input variant performs better on the highest-frequency bin, it exhibits a clear bias toward popular items, whereas AsymRec~maintains a more balanced performance profile across the long-tail distribution.

To further verify the effectiveness of the proposed KRB design, 
we replace KRB with a strong MLP projector as an additional baseline. 
Specifically, we search over the number of layers $\{1,2,3\}$ and hidden dimensions 
$\{64,128,256,512,768,1024\}$, and select the best-performing configuration: 
a 2-layer MLP with dimensions $3072 \rightarrow 512 \rightarrow 448$. 
The resulting model is reported in Row~3 of Table~\ref{tab:ablation}.

As shown in Table~\ref{tab:ablation}, the MLP-based variant already achieves a 
substantial improvement over the discrete-input baseline, demonstrating the 
benefit of adopting continuous semantic representations as model inputs. 
Nevertheless, KRB further improves upon the MLP projector, indicating that the 
proposed structured projection provides additional advantages beyond simply 
introducing continuous inputs. Moreover, KRB achieves better performance with fewer parameters than the MLP projector, which uses a single linear layer with a Kronecker Correction branch containing only 2,560 parameters, indicating that the improvement stems from the structured Kronecker transformation rather than merely increasing projection capacity.

\begin{figure}[tbp]
  \centering
  \includegraphics[width=\linewidth]{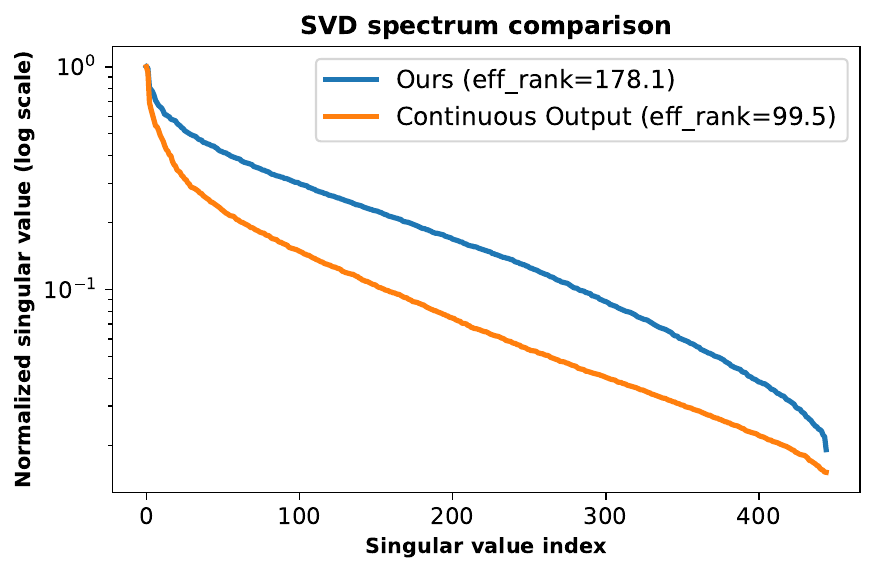}
  \caption{Normalized Singular Spectrum of Transformer Output. We observe that Continuous Embedding Token leads to collapsed singular values, while the Discrete Token leads to more dimensionally robust representations.}
  \label{fig:output_svd_rank}
  \Description{Line plot of normalized singular values for Transformer output representations, showing rapid decay for continuous output and a flatter, more robust spectrum for discrete SID output.}
\end{figure}

\subsubsection{Discrete Output vs. Continuous Output (RQ3)}
\label{sec:continuous_output}

The comparison between Row 1 and Row 4 in Table~\ref{tab:ablation} reveals that adopting a fully continuous pipeline (i.e., predicting continuous embeddings instead of discrete IDs) leads to the most significant performance degradation.

To further investigate the performance discrepancy between continuous and discrete outputs, we hypothesize that the continuous output space suffers from dimensional collapse, a phenomenon where the learned hidden states are restricted to a low-dimensional manifold, thereby limiting the model's expressive capacity. We quantify the dimensionality of the representation space using the Effective Rank (ER)~\cite{roy2007effective}. Specifically, given a matrix of output representations $\mathbf{Z} \in \mathbb{R}^{N \times d}$ for all $N$ predictions, we perform Singular Value Decomposition (SVD) to obtain its singular values $\{\sigma_1, \sigma_2, \dots, \sigma_d\}$. These singular values are normalized into a probability-like distribution $p_i = \sigma_i / \sum_{j=1}^d \sigma_j$, and the Effective Rank is defined as the exponential of the Shannon entropy of this distribution, $ER(\mathbf{Z}) = \exp(-\sum_{i=1}^d p_i \ln p_i)$. As illustrated in Fig.~\ref{fig:output_svd_rank}, the empirical results verify our hypothesis. The normalized singular value spectrum of the continuous variant exhibits a precipitous, power-law decay, where the first few components dominate the variance and the remaining dimensions provide negligible contributions. This leads to a significantly lower effective rank of only 99.5.

In contrast, our AsymRec~ employing discrete SID output maintains a much flatter and more robust singular spectrum. The singular values decay far more gradually, resulting in a substantially higher effective rank of 178.1. This contrast indicates that while direct continuous regression often leads the model toward a "lazy" solution—predicting mean-like vectors that lack discriminative power—supervising the model with discrete classification targets across $M \times L$ subspaces acts as a strong regularizer. By forcing the Transformer to distinguish between diverse semantic clusters defined by MHQ, AsymRec~ effectively prevents the representations from collapsing into a narrow manifold. Consequently, the model preserves a high-dimensional and discriminative feature space, which is essential for capturing the complex, fine-grained item relations required for accurate generative recommendation.

\subsubsection{Impact of MHQ (RQ4)}

\begin{figure}[tbp]
  \centering
  \includegraphics[width=\linewidth]{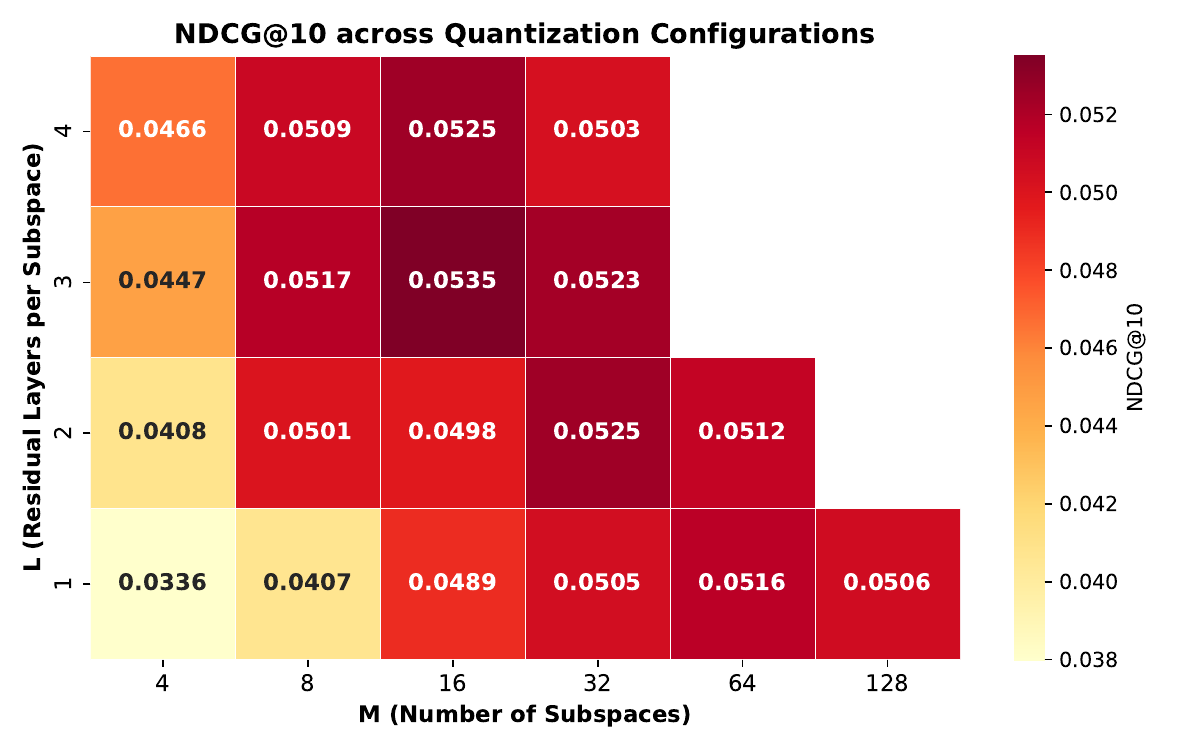}
  \caption{NDCG@10 under different quantization configurations on the Beauty dataset. The horizontal axis shows the number of subspaces $M$, and the vertical axis shows the number of residual layers per subspace $L$.}
  \label{fig:heatmap_M_L}
  \Description{Heatmap of NDCG@10 scores across different numbers of subspaces M (horizontal) and residual layers L (vertical) on the Beauty dataset.}
\end{figure}

Finally, we evaluate the contribution of MHQ by replacing it with a standard PQ approach (Row 5). 
To further investigate the impact of the number of subspaces $M$ and the number of residual layers per subspace $L$, we visualize the corresponding NDCG@10 scores in a heatmap, as shown in Figure~\ref{fig:heatmap_M_L}. 
Here, we only consider configurations where $M \cdot L \le 128$, as increasing $M \cdot L$ beyond this range does not lead to further performance gains.

From the heatmap, it is evident that increasing the number of subspaces $M$ generally improves performance, especially when moving from $M=4$ to $M=64$. 
The effect of adding more residual layers $L$ is more nuanced: moderate increases in $L$ (from $L=1$ to $L=3$) tend to improve NDCG@10, while further increases show diminishing returns. 
This indicates that $M$ and $L$ play complementary roles in capturing the embedding structure. 

Notably, our MHQ design demonstrates significant advantages over standard PQ. 
For example, with $M=8$ and $L=3$, MHQ achieves NDCG@10$=0.0535$ using only 24 tokens, achieving comparable performance to the best PQ configuration ($M=64, L=1$) with 64 tokens (0.0516). 
Overall, these results highlight that MHQ can more efficiently leverage subspace and residual-layer structures to achieve higher recommendation quality with fewer tokens.

\subsection{Online A/B Tests}

To evaluate the practical effectiveness of AsymRec, we conducted online A/B tests within our post-click conversion rate (pCVR) prediction system deployed on one of the world's largest advertising platforms.

In the production environment, AsymRec serves as a semantic ID generation module that supplements the existing ranking pipeline. Item-side continuous semantic embeddings (derived from cross-domain latent factor models and multimodal alignment encoders) are fed into KRB and produce discrete semantic IDs (SIDs). We then map the predicted SID sequence back to a continuous user representation via the corresponding codebook embeddings, and compute the final ranking score via similarity between this user representation and the target item embedding. This design preserves the core asymmetric principle of AsymRec while avoiding the latency overhead of decoding in large-scale serving.

We conducted an online A/B test on a $1\%$ traffic slice over seven consecutive days. Compared to the production baseline, our method achieved a \textbf{1.4\% increase in total consumption} and a \textbf{1.9\% uplift in Gross Merchandise Volume (GMV)}. The improvement demonstrates that the benefits of AsymRec are not limited to offline ranking metrics.

\section{Conclusion}
In this paper, we have identified a critical yet often overlooked limitation in Generative Recommendation: the Dual-stage Information Bottleneck. Symmetric reliance on lossy quantization for both input and output leads to semantic distortion, popularity bias, and imprecise supervision, which collectively hinder the model’s ability to generalize and capture fine-grained item relationships.

To address these challenges, we proposed AsymRec, an asymmetric continuous-discrete framework that decouples the representation paradigms of input and output. On the input side, Kronecker Residual Bridge (KRB) directly maps continuous item embeddings into the model’s feature space, preserving the underlying semantic topology and enabling better generalization to less frequent “cold” items. On the output side, Multi-faceted Hierarchical Quantization (MHQ) constructs high-capacity discrete targets through multi-path, hierarchical quantization, providing precise supervision while preventing dimensional collapse.

Extensive experiments across multiple benchmarks demonstrate that AsymRec consistently outperforms state-of-the-art generative recommenders, validating that asymmetric representation learning—combining continuous input mapping with structured discrete supervision—is key to achieving high-fidelity, fine-grained, and robust generative recommendations.


\bibliographystyle{ACM-Reference-Format}
\bibliography{main}

\appendix
\section{Additional Experimental Results}
\label{sec:appendix}

\subsection{Multi-Dataset Ablation}
\label{sec:appendix_ablation}

To verify that the ablation findings generalize beyond the Beauty dataset, we conduct the key ablations (Row 2: discrete inputs; Row 5: without MHQ) on all four datasets. As shown in Table~\ref{tab:ablation_all}, the conclusions hold consistently: replacing KRB with discrete codes leads to a clear performance drop across all datasets, and removing MHQ similarly degrades performance. These results confirm the robustness of our design choices.

\begin{table}[htbp]
\centering
\caption{Ablation results across all four datasets (NDCG@10).}
\label{tab:ablation_all}
\begin{tabular}{lcccc}
\toprule
\textbf{Variant} & \textbf{Sports} & \textbf{Beauty} & \textbf{Toys} & \textbf{CDs} \\
\midrule
\textbf{AsymRec} & \textbf{0.0311} & \textbf{0.0535} & \textbf{0.0580} & \textbf{0.0511} \\
w/ discrete codes as inputs & 0.0264 & 0.0491 & 0.0506 & 0.0453 \\
w/o MHQ & 0.0290 & 0.0516 & 0.0532 & 0.0493 \\
\bottomrule
\end{tabular}
\end{table}

\subsection{Hyperparameter Sensitivity}
\label{sec:appendix_hyperparam}

\subsubsection{MHQ Regularization Weights}

We evaluate the sensitivity of MHQ training with respect to the loss weights $\lambda_{bal}$ and $\lambda_{reg}$ on the Beauty dataset by varying both parameters in $\{0.001, 0.005, 0.01, 0.05, 0.1\}$. The MHQ training remains stable across all configurations: even the worst-performing setting (NDCG@10 = 0.0524) still outperforms the variant without MHQ (NDCG@10 = 0.0516), demonstrating low sensitivity to these hyperparameters.

\subsubsection{KRB Dimensionality Configuration}

The KRB module involves four dimensionalities: $m_1, m_2$ (satisfying $m_1 m_2 = d = 3072$) and $n_1, n_2$ (satisfying $n_1 n_2 = d_m = 448$). We select these values based on two design principles. First, $m_1$ and $m_2$ should be reasonably balanced: a highly skewed reshape (e.g., $6 \times 512$) reduces the information exchange between the two Kronecker factors $A$ and $B$. Our choice $m_1=48, m_2=64$ (ratio $\approx 2:3$) maintains a nearly square shape. Second, the compression ratios $n_1/m_1$ and $n_2/m_2$ control the structural expressiveness along each axis. We set $n_1=16$ (compression $1/3$) and $n_2=28$ (compression $\approx 0.44$), yielding $2560$ parameters. Empirically, configurations close to this balanced design perform consistently well, while extreme deviations (e.g., $m_1 \ll m_2$) cause the Kronecker correction to degenerate toward the expressiveness of a single linear layer (Beauty N@10 0.0497), losing the benefit over the residual-only baseline.

\end{document}